\pdfoutput=1

\documentclass{article}
\usepackage{spconf,amsmath,graphicx}
\usepackage{siunitx}
\usepackage{multirow}
\usepackage{comment}
\usepackage{diagbox}
\usepackage[subrefformat=parens]{subcaption}

\usepackage{amsmath, amssymb}
\usepackage{type1cm}
\usepackage[legacycolonsymbols]{mathtools}

\title{Noise-robust zero-shot text-to-speech synthesis conditioned on \\self-supervised speech-representation model with adapters}
\name{\begin{tabular}{c}Kenichi Fujita, Hiroshi Sato, Takanori Ashihara, Hiroki Kanagawa\\ Marc Delcroix, Takafumi Moriya, Yusuke Ijima\end{tabular}}

\address{NTT Corporation, Japan}
\begin{document}
\ninept 
\maketitle
\begin{abstract}
The zero-shot text-to-speech (TTS) method, based on speaker embeddings extracted from reference speech using self-supervised learning (SSL) speech representations, can reproduce speaker characteristics very accurately. However, this approach suffers from degradation in speech synthesis quality when the reference speech contains noise. In this paper, we propose a noise-robust zero-shot TTS method. We incorporated adapters into the SSL model, which we fine-tuned with the TTS model using noisy reference speech. In addition, to further improve performance, we adopted a speech enhancement (SE) front-end. With these improvements, our proposed SSL-based zero-shot TTS achieved high-quality speech synthesis with noisy reference speech. Through the objective and subjective evaluations, we confirmed that the proposed method is highly robust to noise in reference speech, and effectively works in combination with SE.
\end{abstract}
\vspace{-1mm}
\vspace{6pt}
\section{Introduction}
\label{sec:intro}
\vspace{-4pt}
Recent text-to-speech (TTS) synthesis trained with a large amount of single-speaker speech data has achieved high-quality natural-sounding human speech~\cite{shen2018natural,ren2020fastspeech}. Furthermore, multi-speaker TTS methods have also demonstrated remarkable results~\cite{chien2021investigating}. These methods facilitate the developments of TTS for any speakers using only a small amount of target speaker utterances, even in a zero-shot setting, by adapting the acoustic model without requiring retraining it~\cite{cooper2020zero,wang2023neural,fujita2023zeroshot}. However, synthesizing high-quality speech from utterances recorded in diverse acoustic environments, e.g., including background noise, remains an ongoing challenge. Overcoming this could expand the utility of zero-shot TTS, e.g., by easing the creation of personalized TTS for patients with speech disorders using noisy recordings, made before patients lose their voice, that are often limited and not of studio-quality.

Most zero-shot TTS methods are based on neural speaker embeddings, i.e., the continuous vector representations of speaker-related information. One popular approach is speaker recognition-based embedding, e.g., d-vector~\cite{heigold2016end,doddipatla2017speaker} and x-vector~\cite{cooper2020zero, snyder2018x}. However, they struggle to capture individual speech rhythms~\cite{fujita21_interspeech}, an important factor among speaker characteristics~\cite{zetterholm2002intonation}. To address this limitation, speaker embedding extraction based on self-supervised learning (SSL) speech model has been proposed~\cite{fujita2023zeroshot}. SSL models, e.g., wav2vec 2.0 (w2v2)~\cite{baevski2020wav2vec}, and HuBERT~\cite{hsu2021hubert}, are trained with a large amount of speech data. As a result, the embeddings extracted from SSL models hold adequate speaker information for zero-shot TTS. Therefore, the SSL-model-based approach is promising for achieving high-quality TTS. 

While some work on noise-robust TTS exists~\cite{zhang2021denoispeech,saeki2022drspeech,nikitaras22_interspeech,yang23i_interspeech}, most of it relies on speaker recognition-based embeddings or speaker-ID based embeddings. NoreSpeech~\cite{yang23i_interspeech} uses a diffusion model to reduce noise in the mel-spectrogram domain to obtain fine-grained style-features and a w2v2 model pre-trained on speaker classification loss for obtaining global embeddings. However, although NoreSpeech aims at zero-shot TTS, this method does not learn robust speaker embeddings optimized for TTS, as its speaker and style encoders are fixed through training. 

To generate robust speaker embeddings for zero-shot TTS, we explore using robust pre-trained SSL models such as WavLM~\cite{chen2022wavlm}. However, WavLM itself may not provide sufficiently robust representation~\cite{hung22_interspeech}. We can fine-tune it, but there are problems, e.g., the computational cost is intensive, fine-tuning is highly time-consuming, and there is a risk of catastrophic forgetting~\cite{french1999catastrophic}. To avoid such problems, we borrow the idea of parameter-efficient tuning (PET) methods. One of the most prevalent PET methods is using adapters~\cite{pmlr-v97-houlsby19a}, that introduce extra tunable weights while keeping the original parameters fixed. This approach has shown comparable performance to the full fine-tuning method~\cite{chen2023chapter}. However, the effectiveness of adapter methods in improving noise-robustness for TTS has not been investigated. 

In this paper, we propose a novel noise-robust zero-shot TTS method based on embeddings extracted from an SSL model. We introduce adapters into the SSL model to adapt the embedding extractor to noisy conditions. By fine-tuning adapters jointly with a TTS model, we can ensure optimal embedding in noisy conditions, while avoiding catastrophic forgetting, which is essential to keep the general speaker representation of the SSL to allow to perform zero-shot TTS. We investigate the effectiveness of adapters for both transformers and a CNN feature extractor, which is a front-end module for SSL models, because recent work~\cite{chen2023chapter} has confirmed that adapters for CNN improve performance in speaker-related tasks, e.g., speaker verification. To further reduce the influence of noise, we use a speech enhancement (SE) fronted before the SSL model, which is shown to be effective at improving noise robustness~\cite{chang22g_interspeech,sato23_interspeech}. Compared with~\cite{yang23i_interspeech}, our proposed method has an advantage in robust speaker embedding extraction as it is optimized for zero-shot TTS because the speaker embedding extraction modules themselves are trained with a TTS model on TTS metrics using noisy speech data. Audio samples are available on our demo page\footnote{https://ntt-hilab-gensp.github.io/icassp2024robustTTS/}. 

\vspace{-6pt}
\section{Proposed Method}
\vspace{-4pt}
We first pre-train a multi-speaker SSL-based TTS model with text-utterance pairs not including noise. Subsequently, we fine-tune the model with text-utterance pairs augmented with noisy audio to improve noise-robustness by only training modules related to speaker embedding extraction, i.e., embedding modules and adapter modules introduced into the SSL model. In this section, we briefly introduce the backbone TTS model and adapters.

\vspace{-4pt}
\subsection{Backbone SSL-based TTS model}
\label{sec:ssl_model}
\vspace{-4pt}

\begin{figure}[tb]
  \centering
  \includegraphics[width=0.7\linewidth]{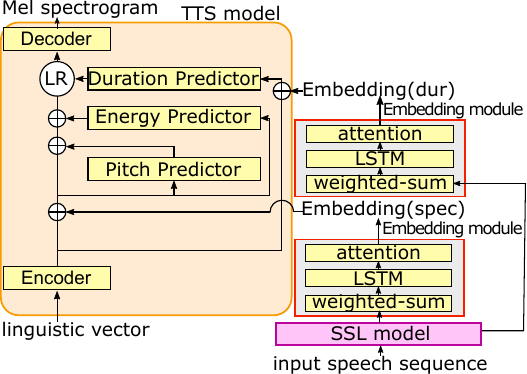}
  \vspace{-2mm}
  \caption{Overview of SSL-based TTS model. Non-autoregressive TTS model is conditioned with latent representations from SSL model. LR denotes length regulator. Duration predictor and other components are conditioned on speaker embeddings separately.}
  \label{fig:overview_sep}
  \vspace{-5mm}  
\end{figure}
Figure~\ref{fig:overview_sep} shows an overview of the backbone SSL-based TTS model~\cite{fujita2023zeroshot}, comprising three components: a non-autoregressive TTS model, e.g., FastSpeech2~\cite{ren2020fastspeech}, an SSL model, and embedding modules. The point is the use of the SSL model for obtaining the embedding vectors and separately conditioning the acoustic features and the duration predictor.

The method uses the SSL model to process an input speech sequence, followed by the embedding module, which converts the speech representations from the SSL model into a fixed length vector, i.e., an embedding vector. This module comprises three parts: weighted-sum, BiLSTM, and attention. In the weighted-sum part, the speech representations from each layer of the SSL model are weighted with learnable weights and summed in the same manner as in~\cite{chen22g_interspeech}. BiLSTM then processes the summed representations, and the hidden states of the BiLSTM are aggregated through an attention~\cite{bhattacharya2017deep, ando2018soft}. Finally, the obtained embedding vectors are fed into the TTS model. Since the TTS model and embedding module are concurrently trained, suitable embedding vectors for the TTS model are obtained from the embedding module.

Furthermore, the acoustic features and the duration predictor are conditioned with separately extracted embeddings~(Fig.~\ref{fig:overview_sep}). This approach enhances speaker modeling precision by obtaining distinct embeddings containing rhythm-based and acoustic-feature-based speaker characteristics, respectively.

\vspace{-4pt}
\subsection{Adapters for SSL model}
\vspace{-4pt}
\begin{figure}[tb]
  \centering
  \begin{minipage}[b]{0.45\linewidth}
    \centering  
    \includegraphics[width=0.7\linewidth]{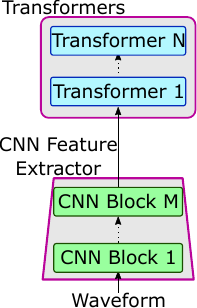}
    \vspace{-0mm}
    \subcaption{SSL model.}
    \label{fig:ssl_model}
  \end{minipage}
  \begin{minipage}[b]{0.45\linewidth}
  \begin{minipage}[b]{1.0\linewidth}
    \centering  
    \includegraphics[width=\linewidth]{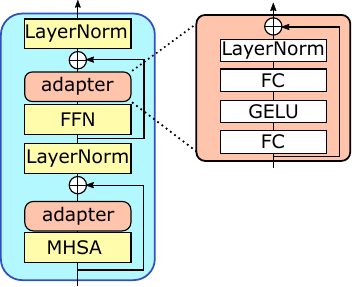}
    \vspace{-6mm}
    \subcaption{Adapter for Transformer.}
    \label{fig:adap_trans}
  \end{minipage}

  \begin{minipage}[b]{1.0\linewidth}
    \centering  
    \includegraphics[width=\linewidth]{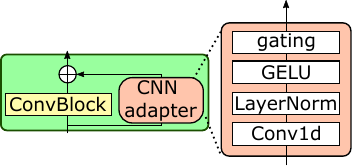}
    \vspace{-6mm}
    \subcaption{Adapter for CNN Block.}
    \label{fig:adap_cnn}
  \end{minipage}
  \end{minipage}
  \vspace{-2mm} 
  \caption{Overview of adapters. Adapters are introduced into transformers and CNN Blocks in SSL model.}
  \label{fig:overview_adapter}
  \vspace{-7mm}  

\end{figure}


\begin{table*}[tb]
  \centering
  \caption{Results of objective evaluations. MCD and Dur. represent mel-cepstral distortion (dB) and RMSE of phoneme duration (ms). FT, BN, and CNN represent fine-tuning, BN and CNN adapters. Values in bold show the best scores in each noise condition.}  
  \label{tab:obeval}
  \vspace*{-3.5mm} 
  \scalebox{0.9}{
  \begin{tabular}{|wc{25mm}|wc{3mm}|wc{3mm}|wc{3mm}|wc{10mm}|c||cc||cc||cc||cc|}
  \hline
  \multirow{3}{*}{method}   &  \multicolumn{3}{c|}{\multirow{1}{*}{conditions}}&  \multirow{3}{*}{\begin{tabular}{c}trainable\\params\end{tabular}} &  \multirow{3}{*}{SNR level} & \multicolumn{4}{c||}{without SE}   & \multicolumn{4}{c|}{with SE}     \\\cline{2-4} \cline{7-14} 
                            &\multirow{2}{*}{FT}&\multirow{2}{*}{BN}&\multirow{2}{*}{CNN}&&               & \multicolumn{2}{c||}{parallel}&\multicolumn{2}{c||}{non-parallel}&\multicolumn{2}{c||}{parallel}&\multicolumn{2}{c|}{non-parallel}\\\cline{7-14} 
                            &&&&&               & MCD  & Dur. & MCD  & Dur.   & MCD  & Dur. & MCD  & Dur.\\ \hline\hline
  \multirow{5}{*}{Pretrained}&\multirow{5}{*}{}&\multirow{5}{*}{}&\multirow{5}{*}{}&\multirow{5}{*}{N/A}&clean       & 5.19 & \textbf{14.8} & 5.92 & 19.8 & 5.21 & 14.8 & 5.94 & 19.8\\
                            &&&&&\SI{20}{dB}          & 5.53 & 15.4 & 6.14 & 20.3 & 5.32 & 15.1 & 6.02 & 19.8\\
                            &&&&&\SI{10}{dB}          & 5.96 & 16.7 & 6.51 & 21.7 & 5.43 & 15.4 & 6.09 & 19.8\\
                            &&&&&\SI{0}{dB}           & 6.90 & 20.4 & 7.24 & 24.6 & 5.63 & 16.8 & 6.22 & 19.9\\
                            &&&&&\SI{-5}{dB}          & 7.53 & 23.8 & 7.72 & 27.3 & 5.78 & 18.1 & 6.32 & 20.6\\ \hline 
  \multirow{5}{*}{Whole FT} &\multirow{5}{*}{$\checkmark$}&\multirow{5}{*}{}&\multirow{5}{*}{}&\multirow{5}{*}{97.2M}&clean      & 5.18 & \textbf{14.8} & \textbf{5.88} & 19.9 & 5.18 & 14.8 & 5.91 & 20.1\\
                            &&&&&\SI{20}{dB}          & 5.26 & \textbf{15.0} & \textbf{5.94} & 19.9 & 5.24 & \textbf{14.9} & 5.94 & 20.2\\
                            &&&&&\SI{10}{dB}          & 5.32 & 15.4 & \textbf{5.99} & 20.0 & 5.29 & 15.2 & 5.96 & 20.1\\
                            &&&&&\SI{0}{dB}           & 5.48 & 16.4 & 6.10 & 20.7 & 5.41 & 15.9 & 6.04 & 20.2\\
                            &&&&&\SI{-5}{dB}          & 5.64 & 17.4 & 6.20 & 21.1 & 5.51 & 16.7 & 6.09 & \textbf{19.8}\\ \hline 
  \multirow{5}{*}{No adapters} &\multirow{5}{*}{$\checkmark$}&\multirow{5}{*}{}&\multirow{5}{*}{}&\multirow{5}{*}{2.89M}&clean       & 5.25 & 15.1 & 5.92 & \textbf{19.5} & 5.24 & 14.9 & 5.94 & 19.7\\
                            &&&&&\SI{20}{dB}          & 5.32 & 15.5 & \textbf{5.94} & \textbf{19.6} & 5.30 & 15.0 & 5.99 & \textbf{19.6}\\
                            &&&&&\SI{10}{dB}          & 5.41 & 16.1 & 6.06 & 20.9 & 5.36 & 15.2 & 6.02 & \textbf{19.6}\\
                            &&&&&\SI{0}{dB}           & 5.66 & 18.2 & 6.30 & 22.2 & 5.50 & 16.2 & 6.11 & \textbf{19.8}\\
                            &&&&&\SI{-5}{dB}          & 5.90 & 19.7 & 6.51 & 23.3 & 5.64 & 17.3 & 6.16 & 19.9\\ \hline 
  \multirow{5}{*}{\begin{tabular}{c}BN adapters\end{tabular}} &\multirow{5}{*}{$\checkmark$}&\multirow{5}{*}{$\checkmark$}&\multirow{5}{*}{}&\multirow{5}{*}{12.4M}&clean      & 5.20 & 14.9 & 5.91 & 19.8 & 5.19 & 14.8 & 5.92 & 20.0\\
                            &&&&&\SI{20}{dB}          & 5.28 & 15.2 & 5.95 & 19.9 & 5.24 & \textbf{14.9} & 5.96 & 19.8\\
                            &&&&&\SI{10}{dB}          & 5.37 & 15.6 & \textbf{5.99} & \textbf{19.9} & 5.29 & \textbf{15.1} & 5.97 & 19.7\\
                            &&&&&\SI{0}{dB}           & 5.57 & 16.6 & 6.10 & \textbf{19.6} & 5.41 & \textbf{15.8} & 6.05 & 20.2\\
                            &&&&&\SI{-5}{dB}          & 5.76 & 17.6 & 6.19 & \textbf{19.7} & 5.52 & 16.7 & 6.07 & 20.0\\ \hline 
  \multirow{5}{*}{\begin{tabular}{c}CNN adapters\end{tabular}} &\multirow{5}{*}{$\checkmark$}&\multirow{5}{*}{}&\multirow{5}{*}{$\checkmark$}&\multirow{5}{*}{7.10M}&clean      & 5.29 & 16.0 & 5.93 & 19.8 & 5.28 & 15.8 & 5.96 & \textbf{19.6}\\
                            &&&&&\SI{20}{dB}          & 5.37 & 16.4 & 5.97 & 19.9 & 5.35 & 16.0 & 6.00 & 19.8\\
                            &&&&&\SI{10}{dB}          & 5.47 & 17.1 & 6.04 & 20.3 & 5.40 & 16.2 & 6.02 & \textbf{19.6}\\
                            &&&&&\SI{0}{dB}           & 5.70 & 18.5 & 6.22 & 21.1 & 5.53 & 17.0 & 6.10 & 20.2\\
                            &&&&&\SI{-5}{dB}          & 5.91 & 19.6 & 6.34 & 21.7 & 5.66 & 18.1 & 6.17 & 20.3\\ \hline 
  \multirow{5}{*}{\begin{tabular}{c}BN and CNN\\adapters\end{tabular}} &\multirow{5}{*}{$\checkmark$}&\multirow{5}{*}{$\checkmark$}&\multirow{5}{*}{$\checkmark$}&\multirow{5}{*}{16.6M}&clean       & \textbf{5.16} & \textbf{14.8} & 5.90 & 19.7 & \textbf{5.17} & \textbf{14.7} & \textbf{5.90} & 19.8\\
                            &&&&&\SI{20}{dB}          & \textbf{5.23} & \textbf{15.0} & 5.95 & \textbf{19.6} & \textbf{5.22} & \textbf{14.9} & \textbf{5.93} & 19.8\\
                            &&&&&\SI{10}{dB}          & \textbf{5.30} & \textbf{15.3} & \textbf{5.99} & \textbf{19.9} & \textbf{5.28} & \textbf{15.1} & \textbf{5.95} & 19.8\\
                            &&&&&\SI{0}{dB}           & \textbf{5.45} & \textbf{16.1} & \textbf{6.09} & 19.8 & \textbf{5.39} & \textbf{15.8} & \textbf{6.03} & 20.0\\
                            &&&&&\SI{-5}{dB}          & \textbf{5.59} & \textbf{17.0} & \textbf{6.16} & 20.0 & \textbf{5.49} & \textbf{16.6} & \textbf{6.04} & 19.9\\ \hline 
  \end{tabular}
  }
  \vspace*{-6.0mm}  
\end{table*}

Figure~\ref{fig:overview_adapter} shows an overview of the adapters. Two types of adapters were introduced to the SSL model: a bottleneck (BN) adapter and CNN adapter. BN adapters are small BN modules inserted between layers of a pre-trained transformer network~\cite{chen22_exploring}. The adapter module consists of two feed-forward layers with a layer normalization and a residual connection. The adapters are inserted after feed-forward networks (FFN) and multi-head self-attention (MHSA) layers in transformers. For stable training, they are initialized with a near-identity initialization. 

We further introduced adapters to the CNN Blocks of the CNN feature extractor similar to CHAPTER~\cite{chen2023chapter}. Our prior work found that the SSL-based TTS method places a relatively large weight on the CNN feature extractor output during the weighted-sum process in the embedding module~\cite{fujita2023zeroshot}. Therefore, fine-tuning with adapters for CNN Blocks would enhance noise-robustness. The CNN adapter module consists of Conv1d with a layer normalization, and a gate structure. These adapters are residually connected to the ConvBlocks. For stable training, a tanh-gating mechanism~\cite{alayrac2022flamingo} is incorporated along with near-identity initialization. This mechanism multiplies the adapter's output by $\mathrm{tanh}(\alpha)$ before adding it to the ConvBlock output, where $\alpha$ is a layer-specific learnable scalar parameter initialized to zero.

\vspace{-6pt}
\section{Experimental Setup}
\vspace{-4pt}
\subsection{Dataset}
\vspace{-4pt}
For the pre-training of the backbone SSL-based TTS model, we used an in-house Japanese speech database which includes 1,071 speakers. This database consists of several speaker types including professional speakers. The database was split into three parts: 147,854 utterances by 967 speakers, 6,743 by 51, and 6,420 by 53 for training, validation, and testing, respectively. The sampling frequency of the speech was \SI{22}{\kHz}. All speech samples were manually annotated with accentual information. 

For the fine-tuning of the pre-trained TTS model, we used speech samples augmented with the MUSAN corpus~\cite{snyder2015musan}. The MUSAN corpus is a music, speech, and noise corpus comprising 109 hours of audio data. The database was split into three parts: 1,620, 207, and 207 samples for training, validation, and testing, respectively. While fine-tuning, the input speech sequences for the SSL model were augmented with randomly selected audio from the MUSAN training data. Note that the noise was added only to the input speech sequence, not to the target sequence of the TTS model.

\vspace{-4pt}
\subsection{Training conditions}
\vspace{-4pt}
The TTS model was FastSpeech2 implemented on the basis of a previous study~\cite{chien2021investigating} where hidden dimensions were set to 256, and encoder and decoder included four and six layers each. The input and target sequences were a 303-dimensional linguistic vector and 80-dimensional mel-spectrograms with a \SI{10.0}{\ms} frame shift. We used WavLM {\sc BASE+}~\cite{chen2022wavlm} for the SSL model because it is the comparably noise-robust SSL model. Although WavLM is not trained with Japanese database, our past research has confirmed the effect of language dependensy was limited~\cite{fujita2023zeroshot}. WavLM processed the input \SI{16}{kHz} raw audio sequence into 768-dimensional sequences, and the embedding modules converted them into 256-dimensional fixed-length vectors. We used HiFi-GAN~\cite{NEURIPS2020_c5d73680} for waveform generation. In pre-training, the backbone TTS model was trained with 400K steps using the Adam optimizer~\cite{kingma-Adam} following the same learning rate schedule as Vaswani et al.~\cite{NIPS2017_3f5ee243}. WavLM was frozen while pre-training. We call this pretrained TTS model \textit{Pretrained}.
 
In fine-tuning, adapters were introduced into WavLM and only the adapters and embedding modules were trained with 200K steps. We fine-tuned the models with the following conditions to confirm the effectiveness of the adapters: with \textit{no adapters}, where only embedding modules were trained, with \textit{BN adapters}, with \textit{CNN adapters}, and with \textit{BN and CNN adapters}. For memory efficiency, the adapters were shared between acoustic and phoneme duration embeddings. For the learning rate, we searched for the best one from the following conditions: fixed to $10^{-3}$ to $10^{-5}$, and the same learning rate scheduled in pre-training. The probability of noise being added to an input speech sequence was 50\% with randomly decided signal-to-noise ratios (SNR) between \SI{-10} to \SI{20}{dB}. The BN size of the BN adapter was 256. Under the same conditions, we also fine-tuned the model without fixing the parameters of WavLM (\textit{Whole FT}) to compare the adapter-based and whole fine-tuning methods. 

Furthermore, we also trained the model by applying an SE fronted before the SSL model to investigate its potential contributions to enhancing noise robustness. We adopted ConvTasnet~\cite{luo2019conv} as the SE front-end module, which converts noisy raw-waveform audio into enhanced raw-waveforms in a time domain. The model was trained with the same training data for the TTS method and the MUSAN corpus. The noise was added at SNR values randomly sampled from \SI{-10} to \SI{20}{dB}. We set the hyperparameters to N=256, L=20, B=256, R=4, X=8, H=512, and P=3 following the notation in~\cite{luo2019conv}. The SE model was frozen in the TTS model training.

\vspace*{-6pt} 
\section{Result}
\vspace*{-4pt} 
\subsection{Objective evaluation}
\vspace*{-4pt} 
We first conducted an objective evaluation to evaluate the performance of the proposed method under \textit{data-parallel} and \textit{data-non-parallel} conditions, where the text to be synthesized corresponds to the text of the reference speech in the former and the latter does not. Under the \textit{non-parallel} condition, one utterance from each speaker was randomly selected as the reference speech. To evaluate robustness against noise, we prepared noisy reference speech conditions adding noise with several SNRs: \SI{20}, \SI{10}, \SI{0}, and \SI{-5}{dB}, in addition to a clean condition. The noise was sourced from the MUSAN test data. The evaluation metrics were mel-cepstral distortion (MCD) and root mean square error (RMSE) of phoneme durations. To calculate MCD between the generated and test speech with the same time alignment without using dynamic time warping, we generated a log mel-spectrogram using the original phoneme durations extracted from the test speech. The RMSE of the phoneme durations was computed by comparing the original phoneme durations of the test speech with those predicted by the duration predictor. Note that under the \textit{non-parallel} condition, objective metrics were obtained comparing the generated speech with the target speaker's speech sharing identical speech content.

Table~\ref{tab:obeval} lists the results of the objective evaluation. A comparison between the conditions with and without SE showed the superiority of the method with SE. Comparing \textit{Pretrained} and others, a performance improvement was observed for most fine-tuning approaches. Therefore, solely applying SE or a noise-robust SSL model, i.e., WavLM, would be insufficient for noise-robust TTS. In the \textit{parallel} condition, the \textit{BN and CNN adapter} condition exhibited superior performance in both MCD and phoneme duration prediction. The performance declined when only the CNN adapters were used, but combined with BN adapters, the best performance was achieved. Furthermore, this condition showed better performance than \textit{Whole FT}. The reason may be attributed to catastrophic forgetting, the loss of knowledge obtained from a large amount of speech data by updating parameters with relatively smaller amount of TTS training data.

Under the \textit{non-parallel} condition, \textit{BN and CNN adapters} consistently showed superiority. Under specific conditions, other methods showed superiority in phoneme duration prediction. This can be attributed to intra-speaker variation. Even for the same speaker utterances, the speech rhythm of each utterance would be inconsistent. Therefore, the proposed method would not necessarily result in improved performance in some \textit{non-parallel} cases, although it can accurately reproduce the characteristics of reference speech.

\vspace*{-4pt} 
\subsection{Subjective evaluation}~\label{sec:subjective_eval}
We conducted a subject evaluation on naturalness and similarity to confirm the effectiveness of the adapters in improving noise robustness and the impact of noise-induced quality degradation. From the results of the objective evaluation, we compared the models with SE in two fine-tuning conditions with and without adapters, i.e., \textit{BN and CNN adapters} and \textit{No adapters}, with three noise conditions for the reference speech, i.e., clean, a low (\SI{10}{dB}) noise ratio, and a high (\SI{-5}{dB}) one. We also used the \textit{Pretrained} condition with clean reference speech as the baseline. For each method, we synthesized 12 sentences from each of four speakers, i.e., two males and two females in the test data, under the \textit{non-parallel} condition. Six noises were selected from three noise categories in the MUSAN test data, i.e., two from each category: music, speech, and general noise. Fifteen participants rated the naturalness of synthetic speech using a mean opinion score (MOS) on a five-point scale from 5: very natural to 1: very unnatural. Similarity was rated on the basis of the degradation MOS (DMOS) on a five-point scale from 5: very similar to 1: very dissimilar, measuring similarity to ground truth recorded speech from the target speaker.

Table~\ref{table:MOS} shows that naturalness remained without deterioration under the low noise condition. Nevertheless, degradation was apparent under the high noise condition, albeit \textit{BN and CNN adapters} showed a lesser degree of deterioration. In similarity, the \textit{No adapters} showed a degradation in quality as noise levels increased. Conversely, \textit{BN and CNN adapters} did not exhibit degradation under the low noise condition. While there was degradation under the high noise condition, the deterioration was less pronounced compared with \textit{No adapters}. Compared with the clean conditions, \textit{BN and CNN adapters} outperformed the \textit{Pretrained} condition. This implies that fine-tuning enhances similarity. Fine-tuning with the introduction of adapters may help alleviate domain mismatch issues between training data for SSL models and that for TTS.

\begin{table}[tb]
  \caption{Naturalness and similarity scores with 95\% confidential interval. BN+CNN represents condition of \textit{BN and CNN adapters}. Values in bold shows the best scores in each noise conditions.}
  \vspace*{-6.5mm}  
  \label{table:MOS}
  \begin{center}
  \scalebox{0.9}{    
  \begin{tabular}{lccc}
  \hline
  \noalign{\vskip.5mm}
    Model& SNR level& MOS-naturalness & DMOS-similarity\\ 
  \hline
  Pretrained        & clean&   $3.42\pm0.08$           &   $3.32\pm0.07$      \\\hline
  \multirow{3}{*}{No adapters}& clean&$3.48\pm0.09$    &$3.28\pm0.07$\\
                              & \SI{10}{dB}&$3.48\pm0.08$    &$3.16\pm0.06$\\
                              & \SI{-5}{dB} &$2.90\pm0.10$    &$2.18\pm0.08$\\\hline
  \multirow{3}{*}{BN + CNN }& clean&$\mathbf{3.49\pm0.08}$    &$\mathbf{3.41\pm0.07}$\\
                            & \SI{10}{dB} &$\mathbf{3.49\pm0.08}$    &$\mathbf{3.36\pm0.07}$\\
                            & \SI{-5}{dB} &$\mathbf{3.26\pm0.09}$    &$\mathbf{3.03\pm0.07}$\\
  \noalign{\vskip.5mm}
  \hline
  
  \end{tabular}%
  }
  \end{center}
 \vspace*{-10.0mm}    
\end{table}

\begin{figure}[tb]
  \centering
  \begin{minipage}[b]{0.7\linewidth}
    \centering  
    \includegraphics[width=\linewidth]{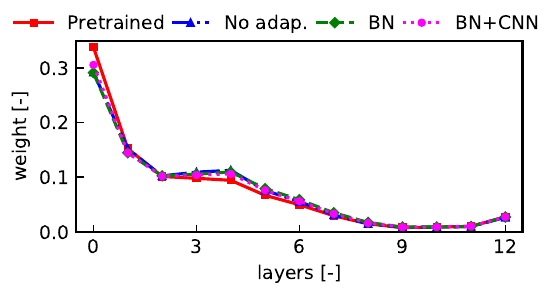}
    \vspace{-6mm}
    \subcaption{Weights that aggregate representations from WavLM for embedding of acoustic features.}
    \label{fig:weight_spec}
    \vspace{-0.5mm}  
  \end{minipage}

  \begin{minipage}[b]{0.7\linewidth}
    \centering
    \hspace{10mm}
    \includegraphics[width=\linewidth]{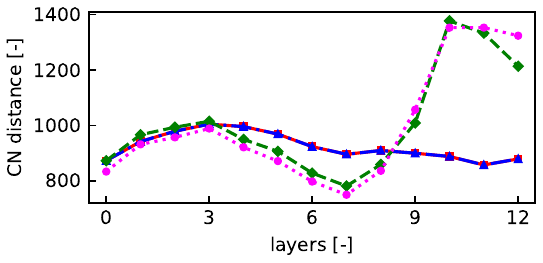}
    \vspace{-6mm}
    \subcaption{Average CN distance.}
  \label{fig:cndist}
    \vspace{-3mm}  
  \end{minipage}
  \caption{Weights for embedding and CN distance. No adap., BN, and BN+CNN represent conditions: \textit{No adapters}, \textit{BN adapters}, and \textit{BN and CNN adapters}, respectively.}
  \label{fig:weight_analysis}
  \vspace{-6.5mm} 
\end{figure}

\vspace*{-5pt} 
\subsection{Noise robustness analysis}
\vspace*{-4pt} 
To analyze the effect of fine-tuning, we analyzed the noise robustness of representations from WavLM. We evaluated the robustness through clean-noisy distance \textit{CN distance}~\cite{hung22_interspeech}, the distance between the latent variables of noisy speech and clean speech defined as $\mathrm{d}_{l}(\mathbf{Z}_c^{(l)}, \mathbf{Z}_n^{(l)}) = 1/T \sum_{t=1}^T \lVert \mathrm{g}^{l}(\mathbf{z}_c^{(l)}(t))- \mathrm{g}^{l}(\mathbf{z}_n^{(l)}(t)) \rVert ^2$, where $l\in \mathbb{N}$ is the layer depth, and $\mathbf{Z}_c^{(l)} \coloneqq {\mathbf{z}_c^{(l)}(t)}_{t=1}^T$ and $\mathbf{Z}_n^{(l)} \coloneqq {\mathbf{z}_n^{(l)}(t)}_{t=1}^T$ denote the set of clean and noisy latent representations for layer $l$ and frame $t$, respectively. $\mathrm{g}^{(l)}$ is a normalization function defined as $\mathrm{g}^{(l)} = (\mathbf{z}^{(l)}(t)-\mathbf{\mu}_{l})/\mathbf{\sigma}_{l}$ with $\mathbf{\mu}_{l}$ and $\mathbf{\sigma}_{l}$ denoting the mean and variance of the latent variable $\mathbf{Z}^{(l)}\coloneqq {\mathbf{z}^{l}(t)}_{t=1}^T$. The smaller \textit{CN distance} implies the effect of noise on representations from WavLM is smaller, i.e., robust to noise.

Figure~\ref{fig:weight_analysis} shows the normalized weights for the embedding module's weighted-sum part and average \textit{CN distance} for the following conditions: \textit{Pretrained}, \textit{No adapters}, \textit{BN adapters}, and \textit{BN and CNN adapters}. Here, the noisy latent representations were extracted from the noisy test speech data at \SI{-5}{dB}, without using the SE front-end. Figure~\ref{fig:cndist} shows that the distance at the 0th-8th layers was comparably small in the \textit{BN and CNN adapters}. Note that, the \textit{CN distance} was particularly small for layer 0, which has the highest weights according to Fig.~\ref{fig:weight_spec}, and thus contributes most to the embedding computation. We consider the CNN adapters contribute to generate representation closer to that of clean speech, i.e., robust to noise, on the 0th layer. Although the distances of the 9th-12th layers were comparably large in \textit{BN and CNN adapters}, the weights for these layers were small; consequently, their impact would be limited.

\vspace*{-8pt} 
\section{Conclusion}
\vspace*{-6pt} 
We proposed a noise-robust zero-shot TTS method conditioned using an SSL model with adapters. Objective and subjective evaluations showed that the proposed method with adapters shows higher robustness for noisy reference speech than a method without adapters. Furthermore, we found that adapters for the CNN feature extractor as well as those for transformers further improve robustness. The effectiveness of the combination of the proposed method and SE was also confirmed. Future work includes applying the proposed method to more sophisticated TTS models such as VITS~\cite{pmlr-v139-kim21f} and JETS~\cite{lim22_interspeech} to achieve higher naturalness and expressiveness.

\newpage
\bibliographystyle{IEEEbib}
{\footnotesize \bibliography{strings,refs}}

\end{document}